\documentclass[%
 reprint,
 amsmath,amssymb,
 aps,
prstab,
]{revtex4-2}
\usepackage{graphicx}% Include figure files
\usepackage{dcolumn}% Align table columns on decimal point
\usepackage{bm}% bold math
\begin{document}

\preprint{APS/123-QED}

\title{BAGELS: A General Method for Minimizing the Rate of \\Radiative Depolarization in Electron Storage Rings}
\thanks{This project was supported by Brookhaven Science Associates, LLC under Contract No. DESC0012704 with the U.S. Department of Energy.}

\author{Matthew G. Signorelli} \email{mgs255@cornell.edu}
\author{Georg H. Hoffstaetter de Torquat} \altaffiliation[Also at ]{Brookhaven National Laboratory}
\affiliation{Department of Physics, Cornell University, Ithaca, NY 14853, USA}

\author{Yunhai Cai}%
\affiliation{SLAC National Accelerator Laboratory, 2575 Sand Hill Road, Menlo Park, California 94205, USA}

\date{\today}% It is always \today, today,

\begin{abstract}
We present a novel method for minimizing the effects of radiative depolarization in electron storage rings by use of vertical orbit bumps in the arcs. Electron polarization is directly characterized by the RMS of the so-called spin orbit coupling function in the bends. In the Electron Storage Ring (ESR) of the Electron-Ion Collider (EIC), as was the case in HERA, this function is excited by the spin rotators. Individual vertical corrector coils in the arcs can have varying impacts on this function globally. In this method, we use a singular value decomposition of the response matrix of the spin-orbit coupling function with each coil to define a minimal number of most effective groups of coils, motivating the name “Best Adjustment Groups for ELectron Spin” (BAGELS) method. The BAGELS method can be used to minimize the depolarizing effects in an ideal lattice, and to obtain fine-tuning knobs to restore the minimization in rings with realistic closed orbit distortions. Furthermore, the least effective groups can instead be chosen for other applications where no impact on polarization is desirable, e.g. global coupling compensation or vertical emittance creation. Application of the BAGELS method has significantly increased the polarization in simulations of the 18 GeV ESR, beyond achievable with conventional methods.%, using only four knobs and minimal orbit excursions.
\end{abstract}

\maketitle

%\tableofcontents

\section{Introduction}
%Synchrotron radiation causes two major effects that dominate the polarization evolution in electron storage rings: (1) the \textit{Sokolov-Ternov effect}, which is an unavoidable asymmetry in the spin flip rate during photon emission leading to a buildup of polarization antiparallel to the bending field, and (2) \textit{radiative depolarization}, which causes a decoherence of the bunch's spins due to the stochasticity of photon emission.
Letting $s$ specify the azimuthal position along the closed orbit and $\langle ... \rangle_s$ denote an average over all phase space coordinates $\vec{r}=(x,p_x,y,p_y,z,\delta)$ at the azimuth $s$, the rate of radiative depolarization in an electron ring can be calculated as

\begin{equation}
    \tau_{dep}^{-1} = \frac{5\sqrt{3}}{8}\frac{r_e\gamma^5\hbar}{m}\frac{1}{C}\oint ds \left\langle g^3\frac{11}{18}\left\lvert\frac{\partial\hat{n}}{\partial\delta}\right\rvert^2\right\rangle_s \ ,
\end{equation}

where $r_e$ is the classical electron radius, $\gamma$ is the Lorentz factor, $C$ is the circumference of the ring, $g=g(s)$ is the local bending strength ($g=1/\rho$ where $\rho$ is the bending radius), $\delta=\frac{\Delta p}{p_0}$ is the relative momentum deviation, and $\hat{n}=\hat{n}(\vec{r},s)$ is the invariant spin field (ISF) \cite{dk}. 
%The Sokolov-Ternov and radiative depolarization effects balance each other out, leading to an asymptotic polarization in the ring defined as
%\begin{equation}
%   P_{\infty} =  \pm\frac{8}{5\sqrt{3}}\frac{\oint ds \ \left\langle g^3\hat{b}\cdot\left(\hat{n}-\frac{\partial\hat{n}}{\partial\delta}\right)\right\rangle_s}{\oint ds \ \left\langle 1-\frac{2}{9}(\hat{n}\cdot\hat{s})^2+\frac{11}{18}\left\lvert\frac{\partial\hat{n}}{\partial\delta}\right\rvert^2\right\rangle_s}\ .
%\end{equation}
The ISF is a special $2\pi$-periodic spin field that solves the Thomas-BMT equation along orbital trajectories,

\begin{subequations} \label{eq:ISF}
\begin{align}
    \hat{n}(\vec{r},s) &= \mathbf{R}(\vec{r}_0,s_0;s)\hat{n}(\vec{r}_0,s_0)\ , \\\hat{n}(\vec{r},s_0 + C)&=\hat{n}(\vec{r},s_0) \ .
    \end{align}
\end{subequations}

On the closed orbit, the ISF reduces to $\hat{n}_0$, which is the 1-turn periodic spin solution on the closed orbit. The projection of a particle's spin along the ISF, or the spin action $J_s=\hat{n}\cdot\vec{S}$, is an adiabatic invariant \cite{georg}. However, in the presence of stochastic radiation, a particle will experience an instantaneous change in its phase space coordinates (kick to $\delta$) when emitting a photon in the bends. The invariant spin direction at its new phase space coordinate will in general be different from the invariant spin direction at its initial phase space coordinate, thus changing the spin action. This effect leads to a rapid decoherence of the particles' spins in an ensemble if left uncorrected \cite{montague}. In a perfectly flat ring with only horizontal dipoles, regular quadrupoles, and regular sextupoles, there are no radiative depolarization effects; the beam will lie almost entirely in the midplane, having an approximately zero vertical beam size due to radiation damping. All magnetic fields in the midplane are vertical in such a ring, and so not only is $\hat{n}_0$ vertical, but so is the ISF over the entire phase space occupied by the beam. Therefore, $\partial\hat{n}/\partial\delta\approx0$. However, once spin rotators are introduced, the regions where $\hat{n}_0$ is not vertical will excite radiative depolarization, because particles' spins will precess different amounts than $\hat{n}_0$ does depending on the phase space coordinates. This causes $\partial\hat{n}/\partial\delta\neq0$.

Traditionally to minimize the effects of radiative polarization, the following procedure is executed \cite{deshandbook}: first, the magnet strengths and spin rotator configuration are set so that, at as many azimuthal positions as possible, the spin dependence after 1-turn on the horizontal, vertical, and longitudinal positions/momenta are removed; this is setting the $\mathbf{G}$-matrices in the SLIM formalism equal to zero \cite{chao}, or equivalently the spin response functions in the spin response formalism equal to zero \cite{sresponse}, and is called \textit{strong synchro-beta spin matching}. Then in a realistic ring, the random closed orbit distortions will cause a tilt of $\hat{n}_0$ in the arcs outside of the spin rotator. This tilt, which will precess with frequency $2\pi a\gamma_0$, will excite $\partial\hat{n}/\partial\delta$. \textit{Harmonic closed orbit spin matching} is performed to correct the tilt by using vertical orbit bumps to cancel out integer spin harmonics nearest to $a\gamma_0$ \cite{Barber:1985nd}. By cancelling out these harmonics, the tilt is reduced. This was used in HERA with success \cite{Barber:1993ui}. In summary, the traditional procedure focuses on reducing the spin-orbit coupling and the integer spin harmonics, which we then expect to also reduce the RMS $\partial\hat{n}/\partial\delta$ globally. 

Here we present a new method, which we call the ``Best Adjustment Groups for ELectron Spin" (BAGELS) method, that instead focuses directly on the RMS $\partial\hat{n}/\partial\delta$ itself in the bends. The BAGELS method is executed by first calculating the response of $\partial\hat{n}/\partial\delta$ at each bend in the ring for each possible ``unit vertical closed orbit bump". Then, a singular value decomposition of the response matrix is performed to obtain a minimal number of best ``groups", or knobs, of corrector coils for electron spin. The BAGELS method can be used for both spin matching the ideal lattice (traditionally performed with strong synchro-beta spin matching) using only vertical corrector coils, and obtaining fine-tuning knobs to correct the spin match with random closed orbit distortions (traditionally performed with harmonic closed orbit spin matching). Furthermore, BAGELS can be used to alternatively obtain knobs that have \textit{no impact} on polarization, for purposes such as delocalized vertical dispersion and/or coupling creation for vertical emittance creation. In this manuscript, we describe how to execute the BAGELS method in detail, and showcase four results of using BAGELS in simulations of the planned 18 GeV Electron Storage Ring (ESR) of the Electron-Ion Collider (EIC): (1) significantly improving polarization in the ideal 1-colliding interaction region (IR) lattice, and more than doubling the polarization in the 2-colliding IR lattice, only using four knobs and a maximum orbit excursion of $<$1.3 mm, (2) restoring the spin match when including random magnet errors (translational/rotational misalignments, field errors) using only four knobs for 10 error seeds, (3) creating a ``polarization-transparent" delocalized vertical dispersion-creating knob for vertical emittance creation, and (4) creating a ``polarization-transparent" delocalized coupling-creating knob for vertical emittance creation.

%For spin matching, the groups with the largest singular values are used to have the maximal impact on polarization with minimal coil strengths and orbit excursions. For almost all other purposes, the groups with the smallest singular values are used for minimal impact on the polarization. 

\section{The BAGELS Method}

Because $\hat{n}_0$ is vertical in the arcs, horizontal orbit bumps in the arcs (with vertical magnetic fields) will have essentially zero impact on the radiative depolarization. Therefore, vertical orbit bumps are to be used. Let $\theta_i$ represent the strength of some arbitrary, single vertical orbit bump. For example, it could be the strength of both coils in a standard $\pi$-bump or a $3\pi$-bump, etc. We will refer to this bump as a \textit{unit bump}, for reasons that will become apparent shortly. The values of $|\partial\hat{n}/\partial\delta|$ at all $m$ bends in the ring can be written as a function of all $n$ unit bumps in the ring, to first-order, as

\begin{align}
\begin{pmatrix}
    |\partial\hat{n}/\partial\delta|_1 \\
    \vdots\\
     |\partial\hat{n}/\partial\delta|_m
\end{pmatrix} = \mathbf{M}_{\textrm m\times \textrm n}\begin{pmatrix}
    \theta_1 \\ \vdots \\ \theta_n
\end{pmatrix} +\begin{pmatrix}
    |\partial\hat{n}/\partial\delta|_1 \\
    \vdots\\
     |\partial\hat{n}/\partial\delta|_m
\end{pmatrix}_0 \ , \nonumber
\end{align}
\begin{align} \label{eg:dnddelta}
    \mathbf{M}_{\textrm m\times \textrm n}=\begin{pmatrix}
        \frac{\partial(|\partial\hat{n}/\partial\delta|_1)}{\partial\theta_1} & \cdots & \frac{\partial(|\partial\hat{n}/\partial\delta|_1)}{\partial\theta_n} \\
        \vdots & \ddots & \vdots \\
        \frac{\partial(|\partial\hat{n}/\partial\delta|_m)}{\partial\theta_1} & \cdots & \frac{\partial(|\partial\hat{n}/\partial\delta|_m)}{\partial\theta_n}
    \end{pmatrix}
\end{align}

$\mathbf{M}_{\textrm m\times \textrm n}$ is the response matrix of $|\partial\hat{n}/\partial\delta|$ at the ends of each bend, for each unit bump. The spin-orbit coupling function can be evaluated using any method (SLIM, spin response, SITF, etc). The derivatives w.r.t each unit bump can be calculated using automatic differentiation if supported by the software, or finite differencing with an appropriately chosen step size (i.e. individually turning on and off every unit bump and calculating the response). We note that one could alternatively use  $1/L\int ds\  g^3\left\lvert \partial\hat{n}/\partial\delta\right\rvert^2$ evaluated at each bend (with the integral perhaps approximated using trapezoidal rule), to directly observe the response of the contribution to $\tau_{dep}^{-1}$. However, we did not find this to be necessary, and in fact all of the results shown in this manuscript only use the response of the $x$-component of the spin-orbit coupling function at each bend.

For spin matching, we now must determine a minimal number of vectors, or ``groups", or ``knobs", that scale the matrix most significantly, so we can get the most ``bang-for-our-buck" using minimal coil strengths and orbit excursions. If $\mathbf{M}_{\textrm m\times \textrm n}$ were a square matrix, we could just take its eigenvectors with the largest eigenvalues. However, for non-square matrices, we must use singular value decomposition (SVD). The compact SVD factors a non-square real matrix into the form

\begin{equation}
    \mathbf{M}_{\textrm m\times \textrm n} = \mathbf{U}_{\textrm m\times\textrm r}\mathbf{\Sigma}_{\textrm r\times\textrm r}(\mathbf{V}_{\textrm n\times\textrm r})^{\textrm{T}} \ ,\  r=\min{(n,m)}
\end{equation}

where $\mathbf{U}^{\textrm{T}}\mathbf{U}=\mathbf{V}^{\textrm{T}}\mathbf{V}=1$ and $\mathbf{\Sigma}$ is a diagonal matrix with the real non-negative singular values $\sigma_i$ in descending order along the diagonal. The columns of $\mathbf{V}$ and $\mathbf{U}$ contain the right- and left-singular vectors respectively, such that

\begin{equation}
    \mathbf{M}\mathbf{v}_i = \sigma_i\mathbf{u}_i\ , \ \mathbf{M}^T\mathbf{u}_i=\sigma_i\mathbf{v}_i \ .
\end{equation}

For spin matching, we therefore choose a minimal number of right-singular vectors with the largest corresponding singular values as our groups, for a maximal impact on the polarization. For other purposes, we generally choose the right-singular vectors with the smallest singular values, for a minimal impact on polarization. These are our ``knobs", that describe how to scale each corrector coil accordingly for a unit change to the knob to optimally achieve the goal.

We now explicitly lay out the BAGELS method, step-by-step:

\begin{enumerate}
    \item \textbf{Choose a unit bump.} The choice of unit bump should be guided by the goal: e.g. for spin matching, a bump that automatically cancels its own generated vertical dispersion and coupling (if there are sextupoles inbetween the coils) is ideal. For global coupling compensation using vertical orbits in sextupoles, a unit bump that intentionally creates delocalized coupling should be used instead.
    \item \textbf{Calculate the response matrix.} The change in $\partial\hat{n}/\partial\delta$ at each bend for each individual unit bump must be obtained to construct the response matrix in Eq.~\eqref{eg:dnddelta}. This can be calculated with automatic differentiation or finite differencing with an appropriate step size.
    \item \textbf{Perform an SVD of the response matrix to get the Best Adjustment Groups for ELectron Spin.} The ``best" groups will be those vectors with either the largest singular values for maximal impact on polarization, or the smallest singular values for minimal impact on polarization, depending on the goal.
\end{enumerate}

\section{Results}
\subsection{Spin Matching the ESR of the EIC}
The 18 GeV ESR of the EIC notably lacks a longitudinal spin match using traditional strong synchro-beta spin matching, due to unfeasibly large solenoid strengths required to achieve such \cite{Signorelli:2023fqc}. The lack of a longitudinal spin match reduces the asymptotic polarization $P_{dk}$ in the 1-colliding IR (1-IP, with only one spin rotator) lattice from 61\% to 35\%, and in the 2-colliding IR (2-IP, with two spin rotators) lattice from 47\% to 17\%. Without any additional correction, the polarization in the ideal 1-IP lattice marginally satisfies requirements, and the polarization in the ideal 2-IP lattice falls short of the requirements. After application of the BAGELS method, we bring the asymptotic polarization to 66\% in the 1-IP lattice and 58\% in the 2-IP lattice, well-exceeding the polarization requirements.

\begin{figure}[!t]
   \includegraphics*[scale=0.65]{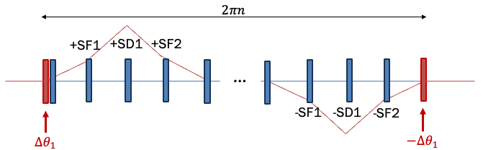}
   \caption{\label{fig:twopin}Schematic of a $2\pi n$-bump that creates localized coupling but delocalized vertical dispersion in arc cells with a 90$^\circ$ degree phase advance and four total sextupole families (two per plane). The blue rectangles represent sextupoles, and the red vertical corrector coils.}
\end{figure}

To apply the BAGELS method, we first choose a unit bump. For spin matching, a unit bump that keeps the generated vertical dispersion and coupling (created by a vertical orbit in sextupoles) localized within the bump is desirable, for minimal impact on the optics. The 18 GeV ESR has a 90$^\circ$ phase advance in the arc cells, with two sextupole families per plane. To localize the coupling in this lattice, we can use a $2\pi n$-bump, as shown in Fig.~\ref{fig:twopin}. With such a bump, all skew-quadrupole terms that the closed orbit sees are cancelled out by an equal but opposite sign skew-quadrupole term in the second half of each oscillation.
To cancel the generated vertical dispersion wave by this bump, we can place a second equal (opposite) $2\pi n$-bump that is out of phase (in phase) with the first. A diagram showing this type of unit bump is shown in the appendix in Fig.~\ref{fig:unit-bump}. We therefore choose our unit bump to be all possible $2\pi n$-bumps with equal (opposite) $2\pi n$-bump out of phase (in phase) in the ring per Fig.~\ref{fig:unit-bump}. Note that we can overlap these bumps with the same dispersion and coupling cancellation, to first-order. 

\begin{figure}[!b]
   \includegraphics*[scale=0.4]{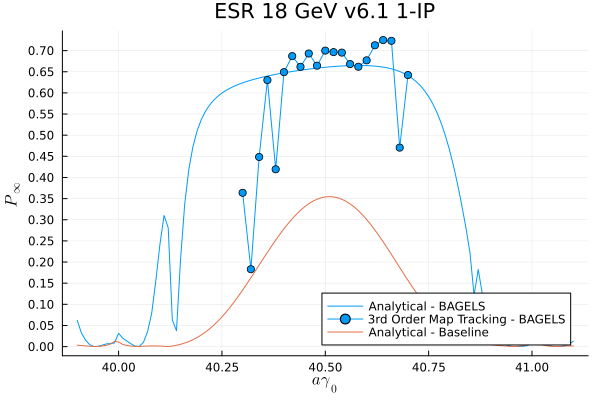}
   \caption{\label{fig:Pdk1}Asymptotic polarization before (orange) and after (blue) applying the BAGELS method for spin matching. 3rd order map tracking refers to calculating $P_{\infty}$ using $\tau_{dep}$ obtained from 3rd order  Monte Carlo tracking with radiation.}
\end{figure}

\begin{figure*}[!t]
   \includegraphics*[width=495pt]{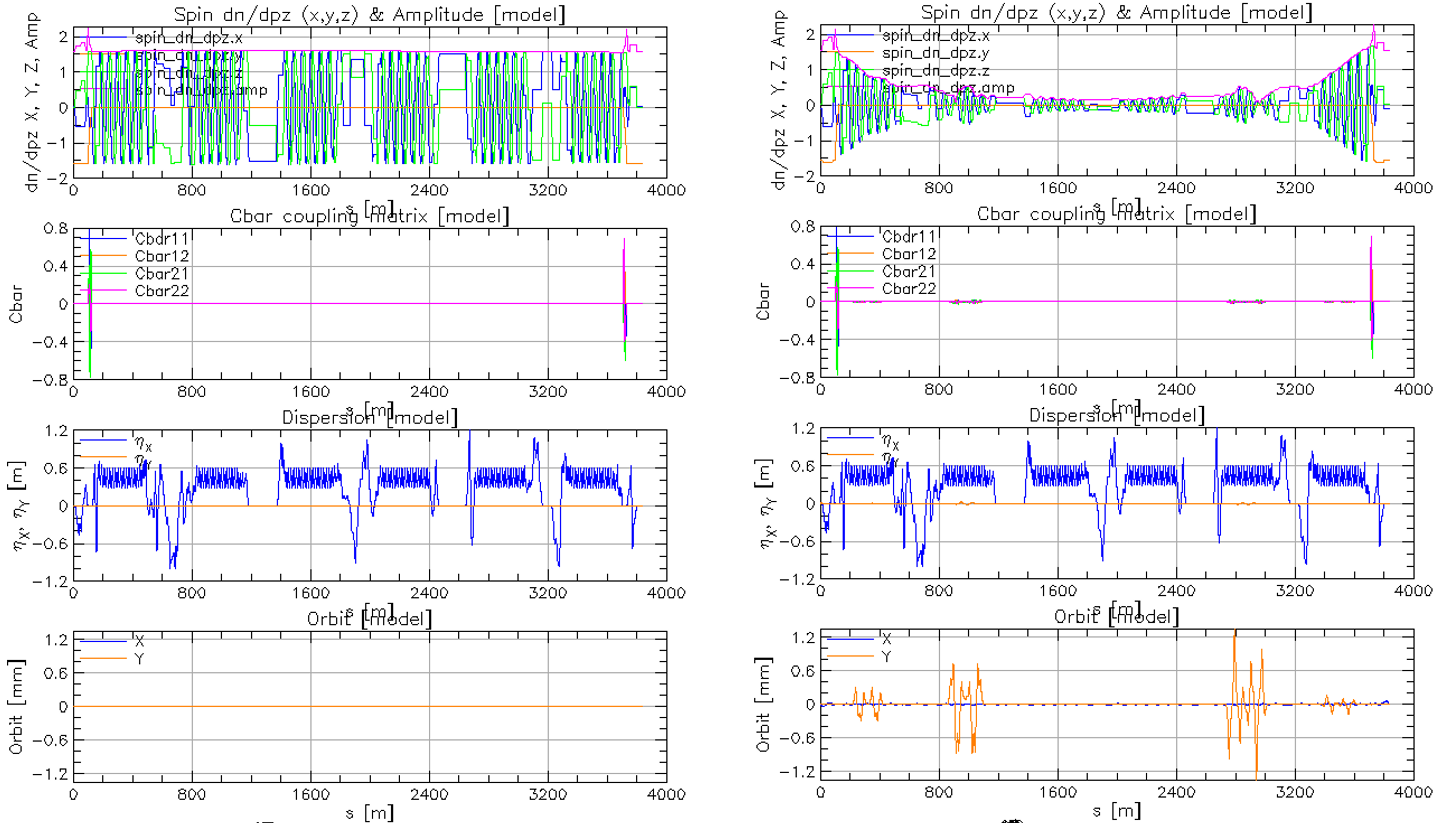}
   \caption{\label{fig:esr1-plot}The 18 GeV 1-IP ESR lattice before (left) and after (right) varying four knobs calculated with the BAGELS method.}
\end{figure*}

After calculating the response of $\partial\hat{n}/\partial\delta$ at each bend for each unit bump, we calculate the SVD of the response matrix and obtain the best groups. Using only the four best knobs, we achieved a significant improvement in the polarization. Figure \ref{fig:Pdk1} shows the difference in the asymptotic polarization $P_{\infty}$ before and after turning the four knobs, obtained with the BAGELS method, in the 1-IP lattice for maximum polarization, and Fig.~\ref{fig:esr1-plot} shows the difference in $\partial\hat{n}/\partial\delta$, dispersion, coupling, and orbit before and after turning the knobs.

\begin{figure*}[!h]
   \includegraphics*[width=495pt]{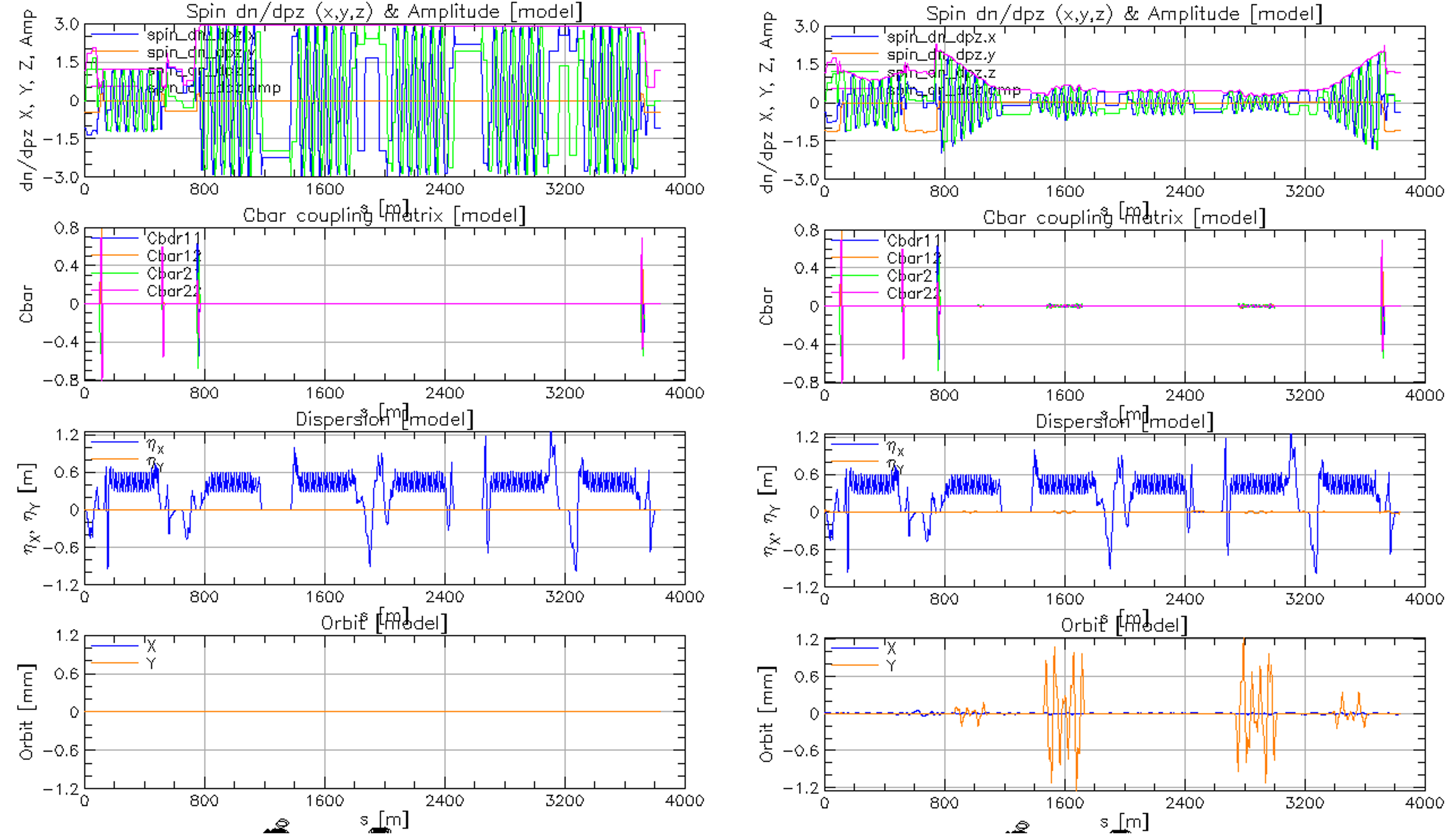}
   \caption{\label{fig:esr2-plot}The 18 GeV 2-IP ESR lattice before (left) and after (right) varying four knobs calculated with the BAGELS method.}
\end{figure*}
\begin{figure}
   \includegraphics*[scale=0.4]{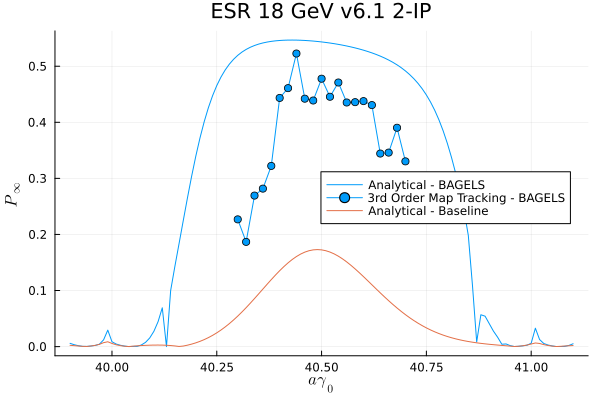}
   \caption{\label{fig:Pdk2}Asymptotic polarization before (orange) and after (blue) applying the BAGELS method for spin matching. 3rd order map tracking refers to calculating $P_{\infty}$ using $\tau_{dep}$ obtained from 3rd order  Monte Carlo tracking with radiation.}
\end{figure}

\clearpage
The polarization requirements are well-exceeded in the 1-IP lattice after application of the BAGELS method, and only four knobs with a maximal 1.3 mm orbit was necessary. Repeating the BAGELS method on the 2-IP ESR, similarly only the best four knobs were needed. Figure \ref{fig:Pdk2} shows the difference in the polarization before and after turning the knobs for maximum polarization for the 2-IP lattice and Fig.~\ref{fig:esr2-plot} shows the difference in $\partial\hat{n}/\partial\delta$, dispersion, coupling, and orbit before and after turning the knobs. The polarization requirements are now exceeded after application of the BAGELS method.

\subsection{Correcting the Spin Match with Random Closed Orbit Distortions}
\begin{figure}[!h]
   \includegraphics*[width=\columnwidth]{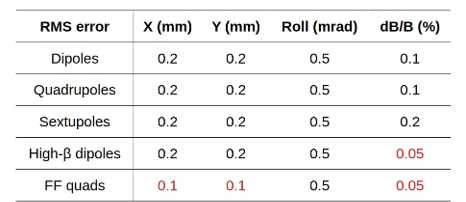}
   \caption{\label{fig:errors}RMS errors used in sensitivity analysis of the 1-IP 18 GeV ESR.}
\end{figure}
In actual operation, there will inevitably be random closed orbit distortions caused by magnet misalignments and field errors. Without any correction, this will degrade the spin match. While traditionally harmonic closed orbit spin matching was used to obtain knobs to restore the spin match, instead the BAGELS method could be used to obtain the best knobs.

\begin{figure}
   \includegraphics*[width=\columnwidth]{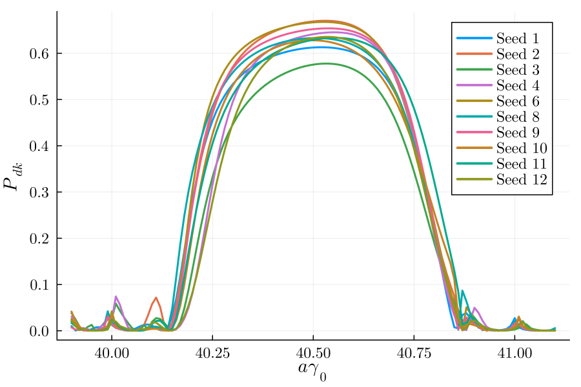}
   \caption{\label{fig:pdkerrors}Asymptotic polarization for 10 different error seeds in the 1-IP ESR after correction of the spin match using five knobs calculated with the BAGELS method.}
\end{figure}

The RMS errors in Fig.~\ref{fig:errors} were used to generate 10 different error seeds of the 1-IP 18 GeV ESR. After correction of the orbit and coupling, five knobs calculated using BAGELS were varied to correct the spin match. Figure \ref{fig:pdkerrors} shows the analytical $P_{\infty}$ obtained after correction of the spin match using the BAGELS knobs for all 10 error seeds.

\subsection{$\epsilon_y$-Creation with Delocalized Vertical Dispersion}
\begin{figure}[!h]
   \includegraphics*[width=\columnwidth]{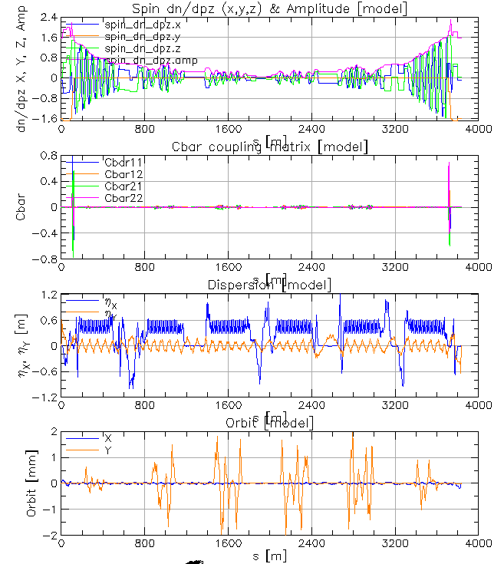}
   \caption{\label{fig:dispersion-tpt}The 18 GeV 1-IP ESR lattice after using ``polarization-transparent" vertical dispersion-creating knobs calculated with the BAGELS method to achieve the desired $\epsilon_y$ for beam size matching.}
\end{figure}

Electron beams in flat rings will have vertical beam sizes very close to zero, due to radiation damping. However, in order to not hurt the proton/light-ion beam lifetime, the electron beam size must be matched to a certain degree with the ion beam at the interaction point(s). One way of achieving beam size matching is to increase the vertical emittance $\epsilon_y$ of the electron beam by generating a delocalized vertical dispersion wave around the ring. However, in general the creation of such a wave will have negative effects on polarization unless done carefully. The BAGELS method can also be applied to create vertical closed orbit bumps that have \textit{no impact} on polarization, by instead choosing the groups to be those with the smallest singular values for minimal impact on polarization.
\begin{figure}[!h]
   \includegraphics*[width=\columnwidth]{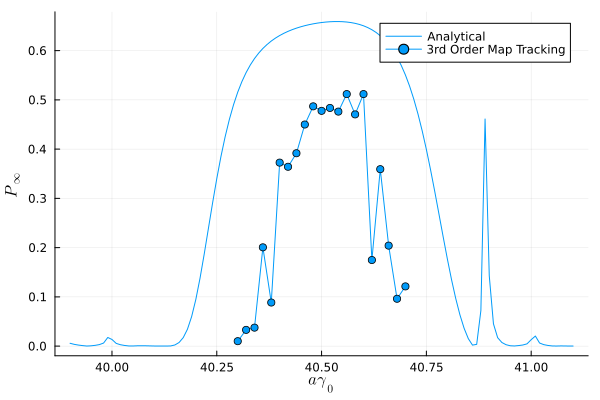}
   \caption{\label{fig:pdkdisp}Asymptotic polarization in the 18 GeV 1-IP ESR after using ``polarization-transparent" vertical dispersion-creating knobs calculated with the BAGELS method to achieve the desired $\epsilon_y$ for beam size matching. 3rd order map tracking refers to calculating $P_{\infty}$ using $\tau_{dep}$ obtained from 3rd order Monte Carlo tracking with radiation. The nonlinear tracking gave an equilibrium $\epsilon_y=$ 3.0 nm vs. the analytical calculation 2.1 nm.}
\end{figure}
Following the procedure, we first choose a unit bump that is best-suited to achieve our goal. In this case, we need to create only delocalized vertical dispersion, without delocalized coupling. Lone $2\pi$-bumps, shown in Fig.~\ref{fig:dispersionbump} give exactly that. Using our BAGELS-spin-matched ESR 1-IP lattice, we reapply the BAGELS method instead using lone $2\pi$-bumps, and now choose our best groups as those with the smallest singular values. These knobs can be turned with essentially zero impact on the polarization, until the desired vertical emittance is obtained. Figure \ref{fig:dispersion-tpt} shows the $\partial\hat{n}/\partial\delta$, dispersion, coupling, and orbit after setting the vertical emittance to 2.1 nm. Comparison of the analytical calculation of $\partial\hat{n}/\partial\delta$ with that in Fig.~\ref{fig:esr1-plot} shows basically no difference, however now the emittance requirements for beam-size matching are satisfied using delocalized vertical dispersion. Figure \ref{fig:pdkdisp} shows a scan of the asymptotic polarization with the desired emittance, with nonlinear tracking. We note that the nonlinearities do cause a greater disagreement in both polarization and the equilibrium vertical emittance (calculated in tracking to be 3.0 nm), however the result is still a significant improvement over the baseline without any vertical emittance creation.

\subsection{$\epsilon_y$-Creation with Delocalized Coupling}
\begin{figure}[h!]
   \includegraphics*[width=\columnwidth]{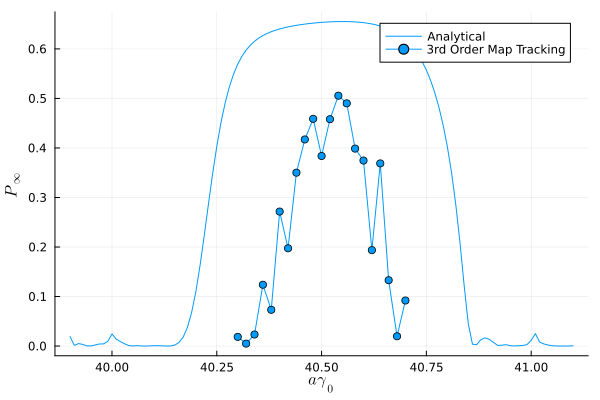}
   \caption{\label{fig:pdkcouple}Asymptotic polarization in the 18 GeV 1-IP ESR after using ``polarization-transparent" coupling-creating knobs calculated with the BAGELS method to achieve the desired $\epsilon_y$ for beam size matching. 3rd order map tracking refers to calculating $P_{\infty}$ using $\tau_{dep}$ obtained from 3rd order Monte Carlo tracking with radiation. The nonlinear tracking gave an equilibrium $\epsilon_y=$ 4.6 nm vs. the analytical calculation 2.0 nm due to a higher-order synchro-beta resonance.}
\end{figure}
\begin{figure}
   \includegraphics*[width=\columnwidth]{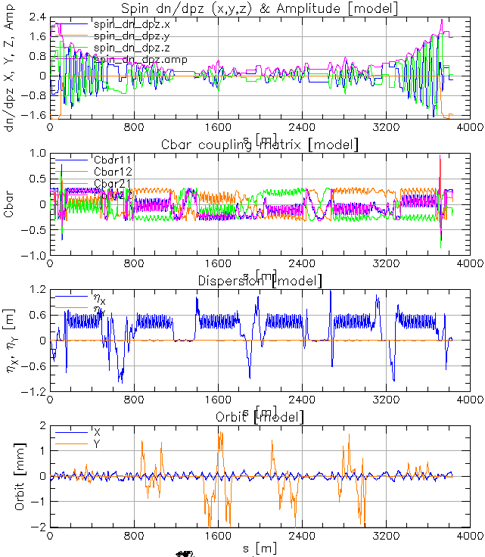}
   \caption{\label{fig:couple-tpt}The 18 GeV 1-IP ESR lattice after using ``polarization-transparent" coupling-creating knobs calculated with the BAGELS method to achieve the desired $\epsilon_y$ for beam size matching.}
\end{figure}

Another way to create vertical emittance is by decoupling some of the horizontal emittance into the vertical. To create only delocalized coupling without any delocalized vertical dispersion in this particular lattice with 90$^\circ$ phase advance and two sextupole families per plane, two equal $\pi$-bumps separated by $2\pi n$ in betatron phase advance can be used, as shown in Fig.~\ref{fig:couplebump}. Applying the BAGELS method using this type of bump as our unit bump, the knobs with the smallest singular values can be varied with minimal impact on the polarization. Figure \ref{fig:couple-tpt} shows the $\partial\hat{n}/\partial\delta$, dispersion, coupling, and orbit after setting the vertical emittance to 2.1 nm. Comparison of the analytical calculation of $\partial\hat{n}/\partial\delta$ with that in Fig.~\ref{fig:esr1-plot} shows very little difference, however now the emittance requirements for beam-size matching are satisfied using delocalized coupling. Figure \ref{fig:pdkcouple} shows a scan of the asymptotic polarization with the desired emittance, with nonlinear tracking. We note that a significant higher-order synchro-beta resonance was observed in nonlinear tracking, suspected to be $Q_y-Q_x-Q_s$, causing the vertical emittance to grow to 4.6 nm. While this is likely the cause for the greater disagreement of the polarization calculated with nonlinear depolarizing effects vs. the linear calculation, the result is still a significant improvement over the baseline even with the very large vertical emittance.

\section{Conclusions}
The Best Adjustment Groups for ELectron Spin (BAGELS) method provides a general approach for obtaining the best knobs of vertical corrector coils for electron spin for a wide range of scenarios. The BAGELS method was used to significantly improve the polarization in both the 1- and 2-colliding IR 18 GeV ESR lattices, in the latter case more than doubling the asymptotic polarizations, with maximum orbit excursions of < 1.3 mm and only four knobs necessary. The BAGELS method was also used to obtain the best knobs to restore the spin match when including random closed orbit distortions with 10 error seeds. Finally, the BAGELS method was used to obtained ``polarization-transparent" knobs that create one of either delocalized vertical dispersion, or delocalized coupling, for vertical emittance creation to achieve beam size matching without any impact on polarization.
\begin{acknowledgements}
We thank Yuri Nosochkov for the suggestion of using a second $2\pi n$-bump to cancel out the dispersion wave created by the first, and Eliana Gianfelice-Wendt for previous work using harmonic bumps to increase polarization in the ESR.
\end{acknowledgements}
\appendix
\section{Unit Bumps}
\begin{figure*}[!t]
   \includegraphics*[width=450pt]{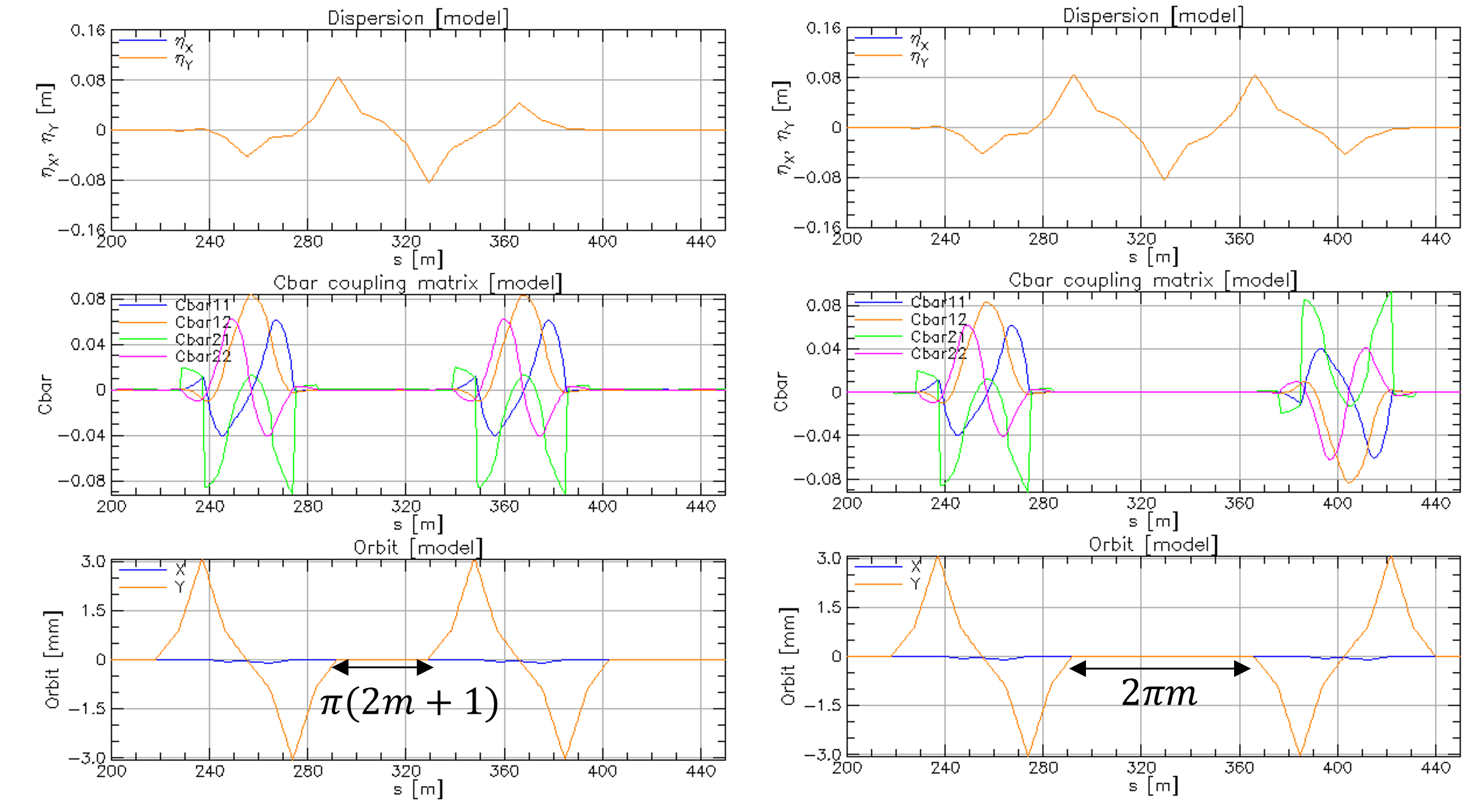}
   \caption{\label{fig:unit-bump} Selected unit bump for spin matching using the BAGELS method, with localized vertical dispersion and localized coupling: (Left) Two equal $2\pi n$-bumps out of betatron phase with each other to cancel the vertical dispersion wave generated by the first. (Right) Two opposite $2\pi n$-bumps in betatron phase with each other to cancel the vertical dispersion wave generated by the first. }
\end{figure*}

\begin{figure}[!b]
   \includegraphics*[width=225pt]{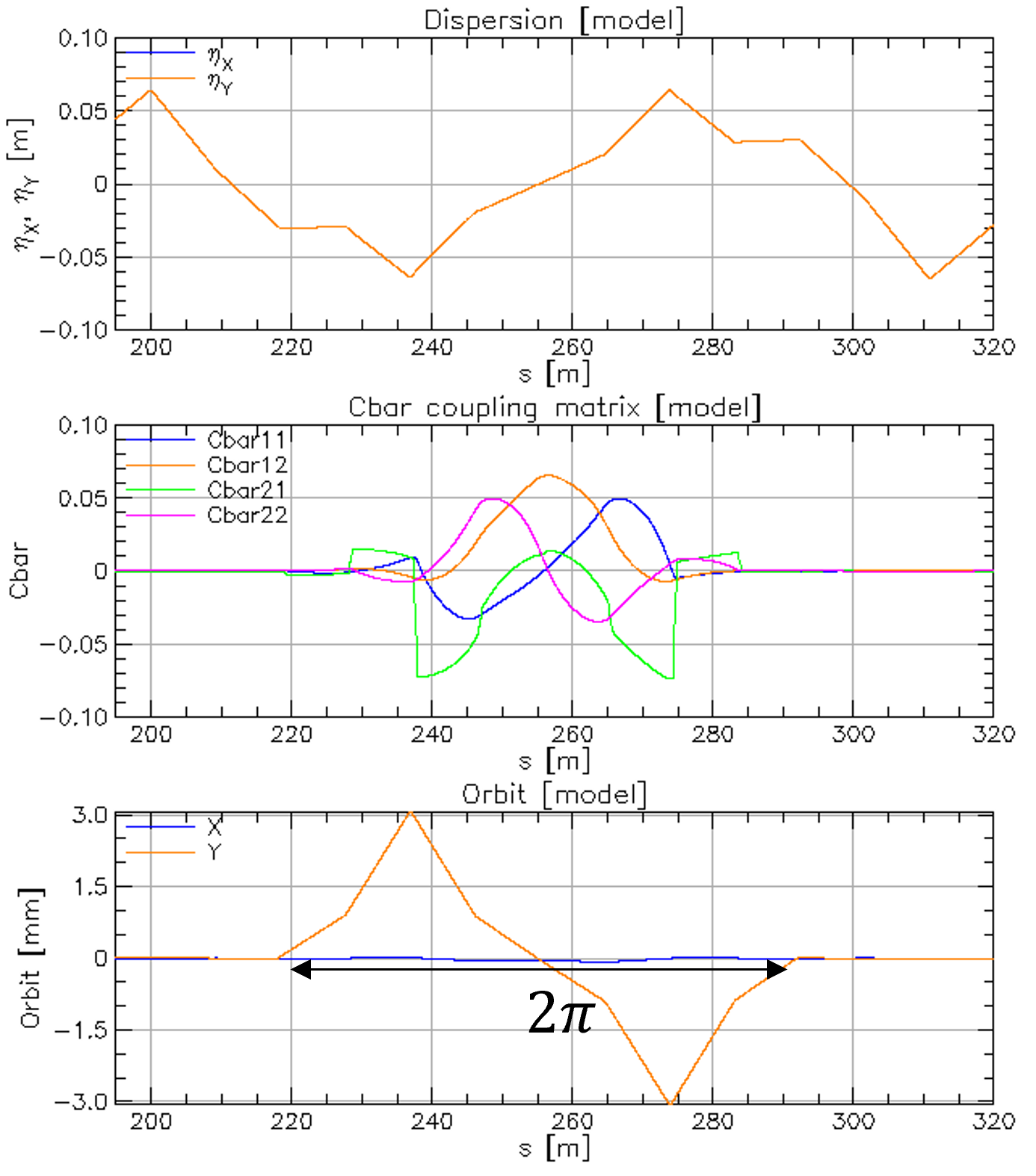}
   \caption{\label{fig:dispersionbump} Selected unit bump ($2\pi$-bump) for creating only delocalized vertical dispersion, with coupling locally cancelled.}
\end{figure}

\begin{figure}[!b]
   \includegraphics*[width=225pt]{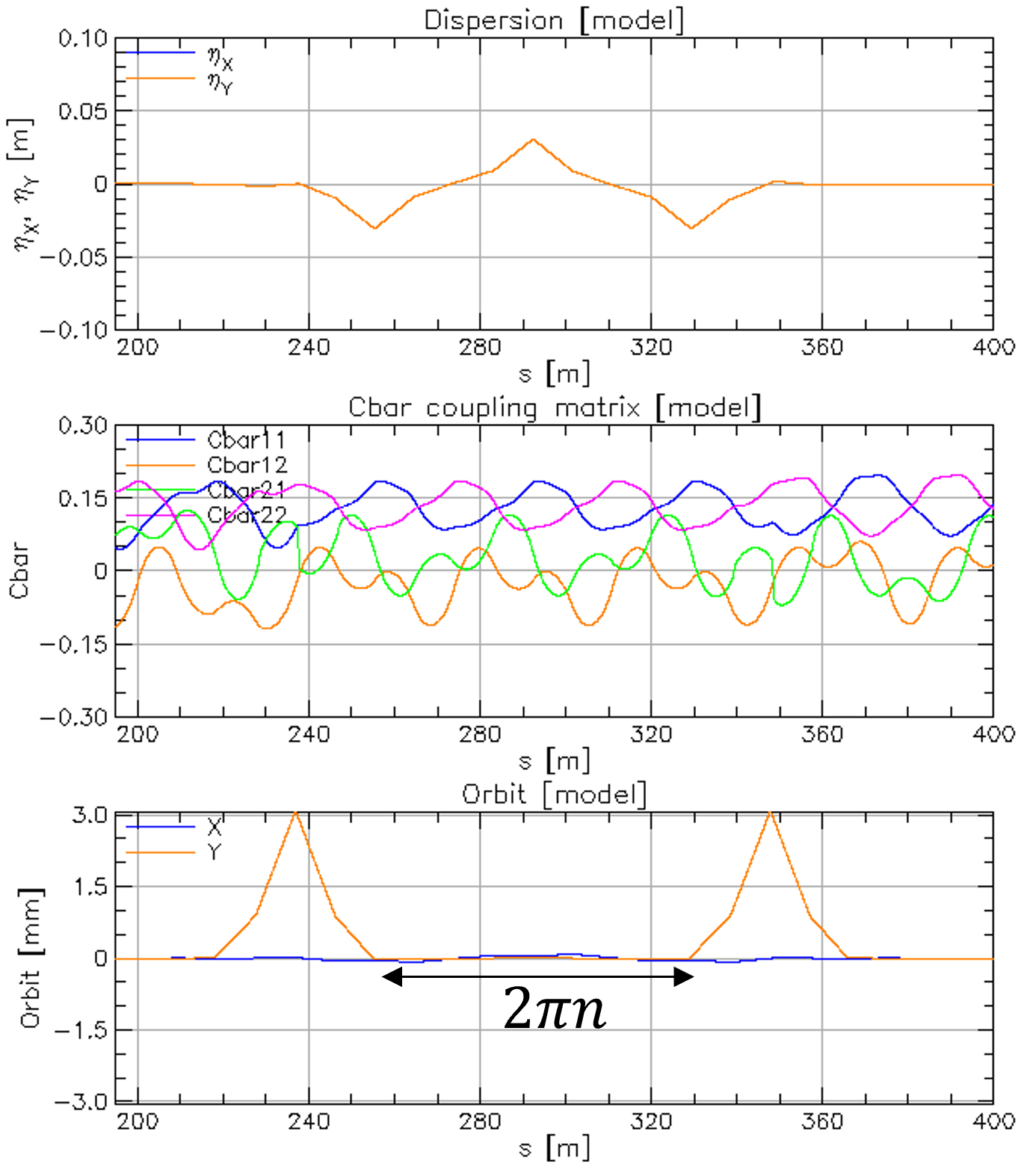}
   \caption{\label{fig:couplebump} Selected unit bump (equal $\pi$-bumps in-phase with each other) for creating only delocalized coupling, with vertical dispersion locally cancelled.}
\end{figure}

\bibliography{apssamp}
\end{document}